\documentclass[prl,twocolumn,superscriptaddress,showpacs,nobalancelastpage]{revtex4}

\usepackage[pdftex]{graphicx}
\usepackage{amsmath}

\newcommand{\SI}[2]{\ensuremath{#1\,\mathrm{#2}}}

\usepackage{xcolor}

\begin{document}

\title{Injection of a single electron from static to moving quantum dots}

\author{Benoit Bertrand}
\affiliation{Univ. Grenoble Alpes, Inst NEEL, F-38042 Grenoble, France}
\affiliation{CNRS, Inst NEEL, F-38042 Grenoble, France}

\author{Sylvain Hermelin}
\affiliation{Univ. Grenoble Alpes, Inst NEEL, F-38042 Grenoble, France}
\affiliation{CNRS, Inst NEEL, F-38042 Grenoble, France}

\author{Pierre-Andr\'e Mortemousque}
\affiliation{Univ. Grenoble Alpes, Inst NEEL, F-38042 Grenoble, France}
\affiliation{CNRS, Inst NEEL, F-38042 Grenoble, France}

\author{Shintaro Takada}
\affiliation{Univ. Grenoble Alpes, Inst NEEL, F-38042 Grenoble, France}
\affiliation{CNRS, Inst NEEL, F-38042 Grenoble, France}
\affiliation{Department of Applied Physics, University of Tokyo, Tokyo, 113-8656, Japan}

\author{Michihisa Yamamoto}
\affiliation{Department of Applied Physics, University of Tokyo, Tokyo, 113-8656, Japan}
\affiliation{PRESTO-JST, Kawaguchi-shi, Saitama 331-0012, Japan.}

\author{Seigo Tarucha}
\affiliation{Department of Applied Physics, University of Tokyo, Tokyo, 113-8656, Japan}
\affiliation{RIKEN Center for Emergent Matter Science (CEMS), 2-1 Hirosawa, Wako-Shi, Saitama 31-0198, Japan}

\author{Arne Ludwig}
\affiliation{Lehrstuhl f\"ur Angewandte Festk\"orperphysik, Ruhr-Universit\"at Bochum, Universit\"atsstrasse 150, 44780 Bochum, Germany}

\author{Andreas D. Wieck}
\affiliation{Lehrstuhl f\"ur Angewandte Festk\"orperphysik, Ruhr-Universit\"at Bochum, Universit\"atsstrasse 150, 44780 Bochum, Germany}

\author{Christopher B\"auerle}
\affiliation{Univ. Grenoble Alpes, Inst NEEL, F-38042 Grenoble, France}
\affiliation{CNRS, Inst NEEL, F-38042 Grenoble, France}

\author{Tristan Meunier}
\affiliation{Univ. Grenoble Alpes, Inst NEEL, F-38042 Grenoble, France}
\affiliation{CNRS, Inst NEEL, F-38042 Grenoble, France}

\date{\today}

\begin{abstract}
We study the injection mechanism of a single electron from a static quantum dot into a moving quantum dot created in a long depleted channel with surface acoustic waves (SAWs). We demonstrate that such a process is characterized by an activation law with a threshold that depends on the SAW amplitude and the dot-channel potential gradient. By increasing sufficiently the SAW modulation amplitude, we can reach a regime where the transfer is unitary and potentially adiabatic. This study points at the relevant regime to use moving dots in quantum information protocols.

\end{abstract}


\maketitle

The ability to displace controllably and on-demand a single electron on a chip is an important prerequisite for the realization of electronic circuits at the single electron level. It opens the route to interconnect nodes of a spin-based quantum nanoprocessor \cite{Taylor2005,Hollenberg2006,Friesen2007, Hu2013} or to perform quantum optics experiments with itinerant electrons \cite{Ji2003,Harju2012,Dubois2013,Bocquillon01032013}. The recent demonstration of fast and on-demand transfer of a single electron is a first step towards this goal \cite{hermelin-nature-2011, mcneil-nature-2011}. In these experiments, the electron is transported in moving quantum dots isolated in an AlGaAs/GaAs heterostructure. They are created in a \SI{4}{\mu m}-long one-dimensional depleted channel by exciting a surface acoustic wave (SAW) \cite{shilton-1996}. Due to the piezoelectric properties of (Al)GaAs, the SAW induces a propagating sinusoidal potential that adds up to the potential of the channel. The resulting moving potential is able to confine and transfer electrons. Such moving quantum dots have permitted to transfer an ensemble of electrons over distances approaching \SI{100}{\mu m} while preserving its spin coherence \cite{ploog:naturemat-2005, santos2013manipulation}. To reach the on-demand single electron transfer regime, an electron is initially isolated in an electrostatic quantum dot located next to the channel entrance, and is then transferred to a moving quantum dot \cite{hermelin-nature-2011}. In order to use such a tool in quantum information protocols, it is of importance to understand the mechanism responsible for the injection of the electron from the static into a moving quantum dot.

In this Letter, we experimentally investigate the injection process from the static to a moving quantum dot. More specifically, we analyzed the influence of two relevant parameters: the SAW modulation amplitude and the static potential induced by the gates defining the source static dot and the channel \cite{kataoka:085302, KataokaSAWPRL2007}. The injection of the electron to a moving quantum dot occurs only when the potential gradient induced by the SAW modulation potential overcomes the one of the static potential. The calibration of the SAW modulation amplitude \cite{hermelin-nature-2011}, and the analysis of the electron injection probability allow us to precisely estimate the value of the static potential gradient for a given gate configuration. This knowledge is used to identify the injection process mechanism. Moreover, it is possible to trigger the electron injection for a nanosecond with a unitary efficiency. We show that this fast injection technique is compatible with an adiabatic injection in which the electron remains in the ground state of the trapping potentials.

\begin{figure}[h!]
\begin{center}
	\includegraphics{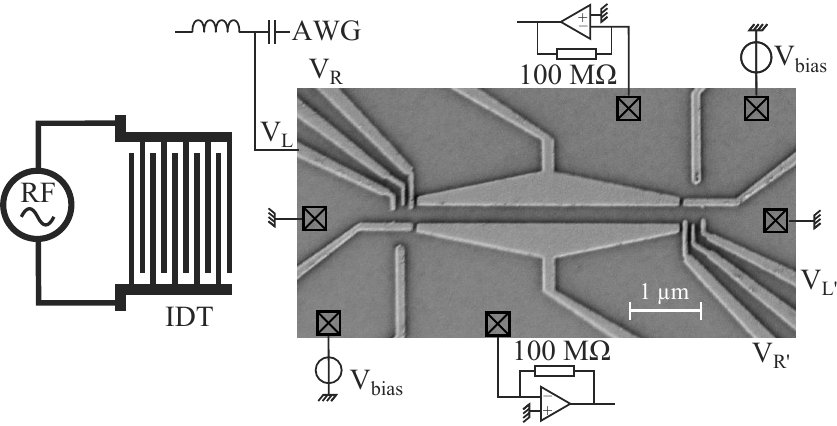}
\end{center}
\caption{SEM picture of a sample similar to the one used for this work, and main electronic set-up. An IDT is placed 2 mm to the left of the structure (schematically sketched on the left of the figure). 
}
\label{fig1}
\end{figure}

The sample used in this work is shown in Fig.~\ref{fig1}. It is fabricated on a GaAs/AlGaAs heterostructure, host of a 2DEG \SI{100}{nm} below the surface, of density $\SI{1.35\times 10^{11}}{cm^{-2}}$ and mobility $\SI{1.5\times 10^{6}}{cm^2\,V^{-1}\,s^{-1}}$. It is constituted of two laterally defined quantum dots, the source and the reception dots, connected together with a $\SI{4}{\mu m}$ long depleted channel. An interdigitated transducer (IDT) is placed $2\,\mathrm{mm}$ to the left of the gate structure in order to generate a SAW. It counts 70 pairs of $60\,\mathrm{\mu m}$ long fingers  with a $1\,\mathrm{\mu m}$ wavelength. When a radiofrequency (RF) excitation of amplitude A$_\text{SAW}$ is applied on the IDT, it resonantly excites a SAW at $2.6323\,\mathrm{GHz}$. Transport measurements through the left (right) quantum dot allowed us to characterize the energy needed for an electron to enter the dot, called the charging energy ($\SI{2.5}{meV}$), and the single-particle orbital energy splitting ($\SI{400}{\mu eV}$). The charge state of the quantum dots is monitored thanks to the adjacent Quantum Point Contacts (QPCs), typically DC-biased with $\SI{300}{\mu eV}$. The sample is anchored to the mixing chamber of a dilution fridge operating at a base temperature of \SI{50}{mK}. 

The experimental sequence that we perform to transfer the electron between the two static quantum dots was initially presented in Ref. \onlinecite{hermelin-nature-2011}. The source dot is isolated from the lead allowing to trap the electron \SI{5}{meV} above the Fermi energy. We then add a $\SI{10}{\mu s}$ voltage pulse on V$_\text{R}$ and V$_\text{L}$, characterized by a voltage displacement ($\delta \text{V}_\text{L}$,  $\delta \text{V}_\text{R}$) called the sending pulse. As a result, the potential gradient between the channel and the source dot is further reduced during the pulse. A few hundred nanosecond RF-excitation is applied on the IDT to generate a SAW modulation within the duration of the sending pulse, ensuring the electron transfer. The presence/absence of the electron can be inferred in both the source dot and the reception dot via charge measurements \cite{field-prl-1993,hermelin-nature-2011}. 

\begin{figure}[h!]
\begin{center}
\includegraphics[width=3.4in]{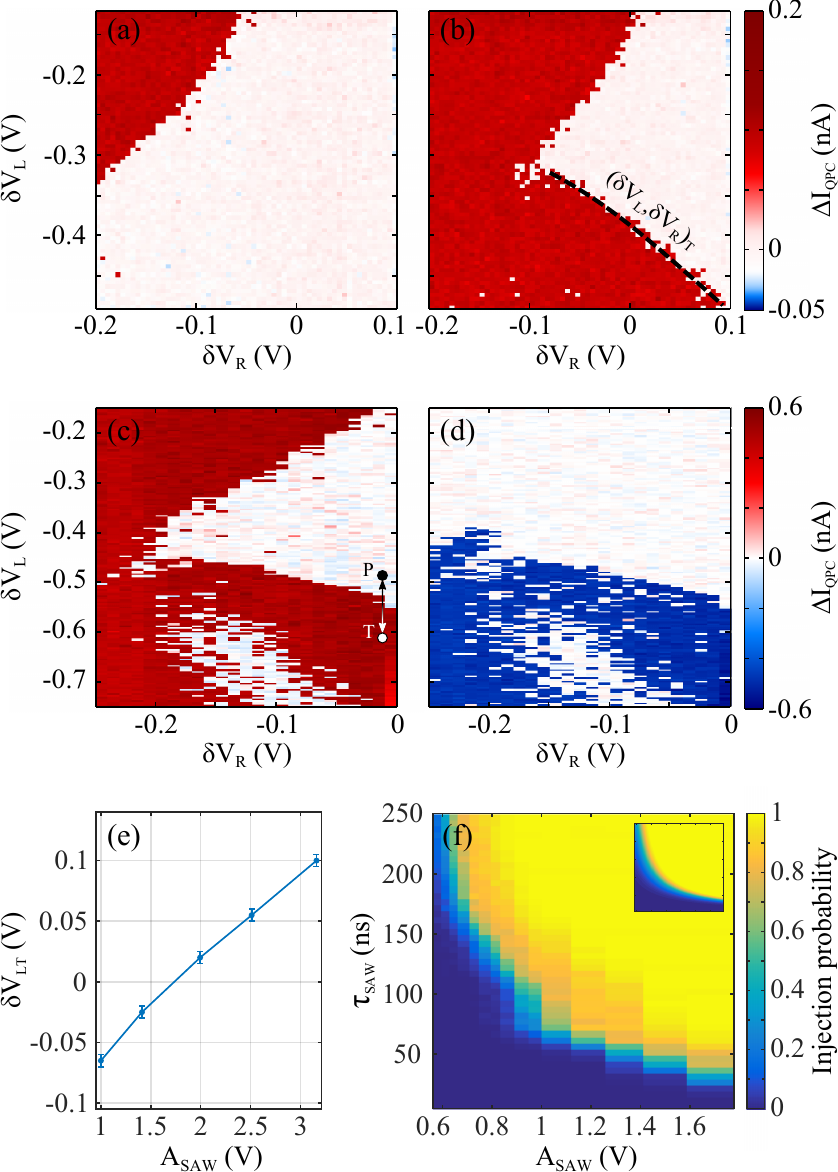}
\end{center}
\caption{(color online) Difference of the QPC current recorded with the charge detector (either on the source or the reception dot) before and after applying a sending pulse ($\delta \text{V}_\text{L}$, $\delta \text{V}_\text{R}$) on the source dot. Each point is the result of a single realization of the experiment. Red and blue points correspond to the respective cases where one electron leaves or is added to the dot after the sending pulse. White points corresponds to no change in the dot charge configuration.  In (a) and (b), we concentrate on the source dot charge configuration, respectively without and with SAW modulation during the sending pulse. (c) and (d) present respectively the QPC response of the source and the reception dots with the SAW modulation applied during the sending pulse. The observed correlations between the two plots confirm that an electron arrives in the reception dot only for the sending pulse corresponding to the transfer region. It is worth noting that the two sets of data (a,b) and (c,d) correspond to slightly different gate voltage configurations. (e) Evolution of the position of the voltage threshold $\delta \text{V}_{\text{L}T}$ as a function of $\text{A}_{\text{SAW}}$ [set of data (a,b)]. (f) Emission probability as a function of $\text{A}_{\text{SAW}}$ and $\tau_{\text{SAW}}$ with a sending pulse ($\delta \text{V}_\text{L}=\SI{-0.55}{V}$, $\delta \text{V}_\text{R}=\SI{0}{V}$) [set of data (c,d)] recorded over 500 experimental realizations. (Inset) Fit of the data with an activation law with a threshold A$_T=\SI{0.98}{V}$ and a typical timescale $\tau_{\text{IDT}}=\SI{63}{ns}$.}
\label{fig2}
\end{figure}

To study the influence of the dot potential on the injection process from the source to the moving dots, we vary ($\delta \text{V}_\text{L}$, $\delta \text{V}_\text{R}$) and probe the charge present in the source dot after the sending pulse [see Fig.~\ref{fig2}(a-c)]. Without SAW modulation [Fig.~\ref{fig2}(a)], the electron remains in the isolated dot (white region) except for the sending pulses with the most positive V$_\text{L}$ and the most negative V$_\text{R}$ (red region): the dot isolation is no longer sufficient and the electron is lost to the neighbor reservoir. With SAW modulation [Fig.~\ref{fig2}(b,c)], a new region where the electron leaves the source dot, called the transfer region, is present for the most negative V$_\text{L}$ and V$_\text{R}$ (bottom left corner). It corresponds to the process where the electron has been injected in a SAW moving quantum dot and leaves the source dot. A confirmation of these two regions labeling is obtained by analyzing the correlation between the electron leaving the source dot [Fig.~\ref{fig2}(c)] and arriving in the reception dot (initially empty) after the transfer sequence [Fig.~\ref{fig2}(d)]. An electron is indeed detected in the reception dot after the transfer only for the sending pulses corresponding to the transfer region.

The sending pulse amplitude has to be more negative than a certain threshold on ($\delta \text{V}_\text{L}$, $\delta \text{V}_\text{R}$) [see Fig.~\ref{fig2}(c)], defined by the limit $(\delta \text{V}_\text{L}, \delta \text{V}_\text{R})_T$, that separates the transfer region and the region where the electron remains in the source dot. In the following, we analyze the behavior of the transfer for $\delta \text{V}_\text{R}=0$. We find that the transfer is activated only below a certain threshold $\delta V_{\text{L}T}$,  that linearly depends on $\text{A}_{\text{SAW}}$ [see Fig.~\ref{fig2}(e)]. Indeed for a larger $\text{A}_{\text{SAW}}$, the potential gradient imposed by the SAW modulation overcomes more easily the one resulting from the gate voltages. It results in a wider range of gate voltages where the injection of the electron to a moving quantum dot is possible. This demonstrates the dependence of the two important parameters for the electron transfer: the potential gradient induced by the SAW, and the static dot potential. We conclude that the electron transfer is activated above a threshold defined by two interdependent parameters A$_T$ and V$_T$.

This model has to be refined by taking into account the risetime of the SAW modulation that is due to the finite bandwidth of the IDT. To investigate precisely this aspect, we set the sending pulse parameters ($\delta \text{V}_\text{L}$, $\delta \text{V}_\text{R}$) to (\SI{0}{V}, \SI{-0.55}{V}) and varied both $\text{A}_{\text{SAW}}$ and the RF-excitation duration $\tau_{\text{SAW}}$. The resulting probability to emit the electron is represented in Fig.~\ref{fig2}(f). The injection of the electron to a moving quantum dot occurs only when $\text{A}_{\text{SAW}}$ reaches the amplitude threshold $\text{A}_\text{T}$. As expected from the risetime of the SAW modulation, a shorter $\tau_{\text{SAW}}$ is needed to reach $\text{A}_T$ when $\text{A}_{\text{SAW}}$ is increased. A fit of the data is performed assuming an activation law [see inset of Fig.~\ref{fig2}(f)]. It enables us to deduce the risetime of the SAW modulation (about \SI{63}{ns}) and the value of $\text{A}_T$ (about \SI{1}{V}). From the relation between $\delta V_{\text{L}T}$ and $\text{A}_{\text{SAW}}$, obtained from Fig.~\ref{fig2}(e), and the gate lever arm ($\approx$ \SI{20}{mV/meV}), we estimate that $\text{A}_T$ corresponds to a SAW-induced potential gradient seen by the electron $\Delta \text{V}_{\text{SAW}}$ $\approx$ \SI{30}{meV/\mu m}. As explained earlier, it corresponds also to the dot-channel potential gradient in the specific gate configuration of Fig.~\ref{fig2}(f).

\begin{figure}[h!]
\begin{center}
\includegraphics[width=3.4in]{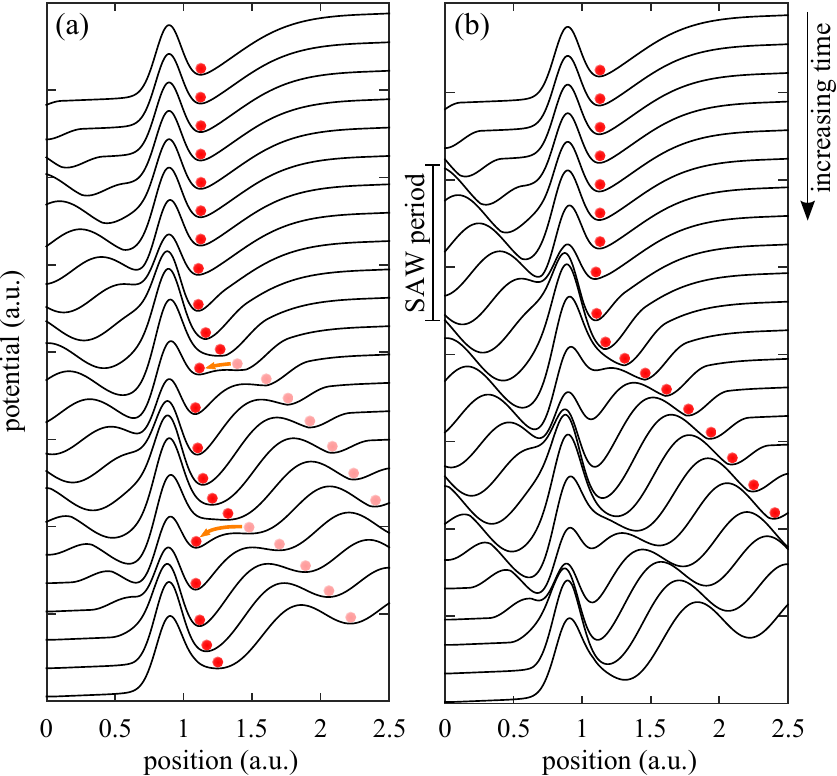}
\end{center}
\caption{(color online) Time evolution of the dot potential for two different amplitudes of SAW excitation. The time increases going from top to bottom traces and the time difference between two consecutive traces is $T/13$, where $T$ is the period of the SAW excitation. The potential gradient $\Delta \text{V}_\text{g}$ induced by the gate voltages is the same in (a) and (b). (a) The potential gradient induced by the SAW excitation $\Delta \text{V}_{\text{SAW}}$ is comparable to $\Delta \text{V}_\text{g}$ and shallow moving quantum dots are defined. Tunnelling back of the electron to the source dot is possible (see orange arrows), leading to a stochastic injection procedure. In (b) $\Delta \text{V}_{\text{SAW}}$ is larger than $\Delta \text{V}_\text{g}$ and deeper moving quantum dots are defined. 
}
\label{fig3}
\end{figure}

To have a better insight into the injection process, we sketched the time dependent potential seen by the electron in the presence of the SAW excitation (see Fig.~\ref{fig3}). At a fixed potential gradient induced by the gate voltages $\Delta \text{V}_\text{g}$, the transfer process depends on $\Delta \text{V}_{\text{SAW}}$. For $\Delta \text{V}_{\text{SAW}}$ lower than $\Delta \text{V}_\text{g}$ [Fig.~\ref{fig3}(a)], shallow SAW moving dots are created and a thin tunneling barrier separates them from the source dot. As a consequence, a process where the electron can tunnel back to the source dot after being caught is likely to happen \cite{kataoka:085302, KataokaSAWPRL2007}. Considering the \SI{30}{meV/\mu m} dot-channel potential gradient previously estimated, the energy detuning between the source and the moving quantum dots in the tunneling-back configuration is expected to be larger than the orbital energy splitting of the source dot. Therefore, the tunneling-back events will certainly result in the excitation of higher orbital states of the source dot, and consequently the injection process will be non-adiabatic in this situation. For $\Delta \text{V}_{\text{SAW}}$ bigger than $\Delta \text{V}_\text{g}$ [Fig.~\ref{fig3}(b)], the tunnel barrier becomes too thick and no tunneling-back process is possible. In this situation, the injection process is expected to be adiabatic and happens when the source and the moving quantum dot are overlapping. 

Therefore, under the experimental conditions of Fig.~\ref{fig2}(f), we infer that the injection process is certainly stochastic and, according to the previous discussion, non-adiabatic. One can notice on Fig.~\ref{fig2}(f) that the electron is expelled from the dot after at least a few tens of nanoseconds corresponding to the time needed for $\text{A}_{\text{SAW}}$ to reach A$_T$. The electron experiences the excitation of a few hundreds of moving quantum dots before being transferred with most part of it at $\text{A}_{\text{SAW}}$ below A$_T$, in a configuration similar to Fig.~\ref{fig3}(a). The emission probability reaches unity only because the electron has the possibility to be injected in a considerable number of SAW periods. It also results in the exploration of the excited states of the source dot and in a non-adiabatic injection process. To be able to probe the adiabatic limit, it is therefore necessary to characterize the electron transfer at a constant SAW modulation amplitude and for a time as short as possible.

To this aim, a modified transfer sequence was utilized for the injection of a single electron. $\tau_{\text{SAW}}$ was reduced to its minimum duration allowed by the SAW risetime, and we added a \SI{1}{ns} pulse on gate $\text{V}_\text{L}$ synchronized with the SAW modulation \citep{hermelin-nature-2011}. This gate pulse is used to rapidly change the potential gradient between the dot and the channel. Practically, the dot potential is tuned to a point P [see Fig.~\ref{fig2}(c)] where the potential gradient is too large for the SAW to transfer the electron into the moving quantum dot. Only for the \SI{1}{ns} pulse, the system is tuned to point T [see Fig.~\ref{fig2}(c)] where the dot-channel potential gradient is reduced. As a consequence the electron is expected to be transported by the SAW only during the nanosecond pulse. This technique allows us to probe the injection process only for two periods of the SAW excitation and more importantly at constant $\text{A}_{\text{SAW}}$.

\begin{figure}[h!]
\begin{center}
\includegraphics[width=3.4in]{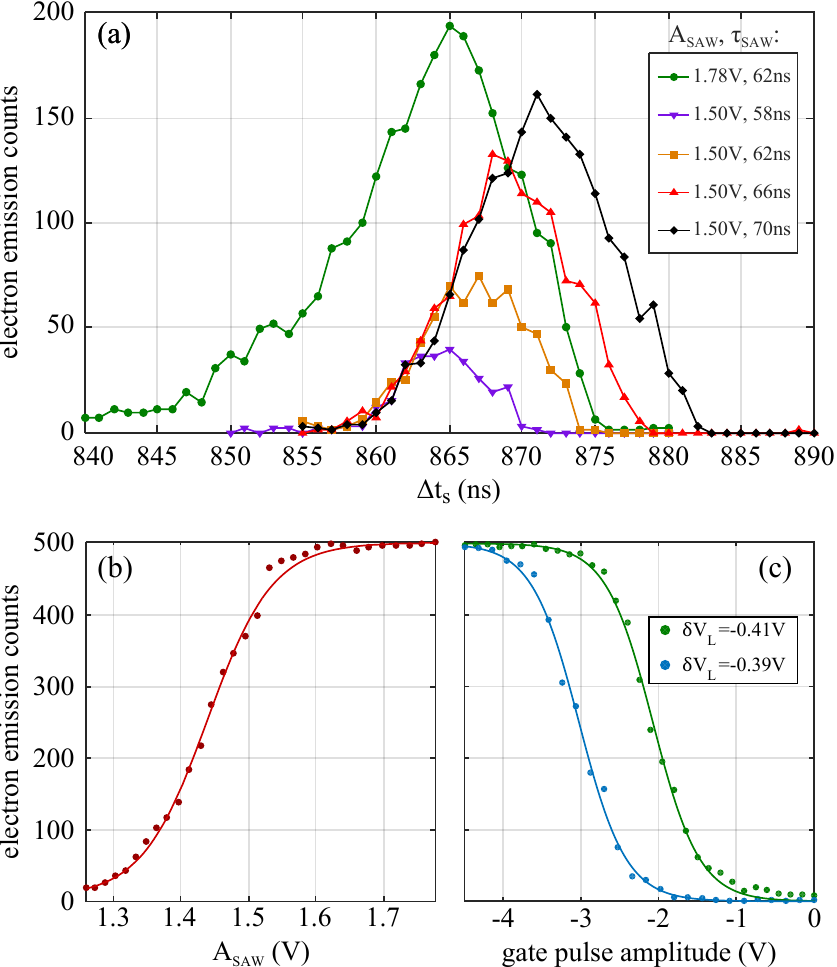}
\end{center}
\caption{(color online)(a) Electron emission counts out of 200 repetitions as a function of the delay between the SAW burst generation and the nanosecond pulse for different $\text{A}_{\text{SAW}}$ and $\tau_{\text{SAW}}$.
(b) and (c) Electron emission counts out of 500 repetitions at $\Delta \text{t}_{\text{max}}$ as a function of the $\text{A}_{\text{SAW}}$ (b) and the gate pulse amplitude (c). $\tau_{\text{SAW}}$ is fixed at \SI{62}{ns} in both cases.
}
\label{fig4}
\end{figure}

Figure~\ref{fig4}(a) shows the injection counts as a function of the delay $\Delta \text{t}_\text{s}$ between the SAW burst and the nanosecond pulse applied to the gate. For $\text{A}_{\text{SAW}}=\SI{1.78}{V}$, the injection probability reaches almost unity for a delay close to $\Delta \text{t}_{\text{max}}=$ \SI{865}{ns}. The delay is explained by the finite SAW velocity of \SI{2867}{m/s} \cite{takagakisantos2002} and the distance between the IDT and the sending dot. Before and after $\Delta \text{t}_{\text{max}}$, $\text{A}_{\text{SAW}}$ is not high enough to ensure a unity transfer process and the electron can be back-scattered to the source dot. An exponential RC rising/falling time law for $\text{A}_{\text{SAW}}$ due to the finite risetime of the SAW excitation explains qualitatively the asymmetric shape of the emission probability as a function of $\Delta \text{t}_{\text{s}}$. $\Delta \text{t}_{\text{max}}$ corresponds therefore to the end of the SAW excitation burst, when the amplitude is maximal. For $\text{A}_{\text{SAW}}=\SI{1.5}{V}$, the injection probability follows the same shape with a lower success rate. When the SAW excitation is lengthened from \SI{58}{ns} to \SI{70}{ns} at \SI{1.5}{V}, the maximum SAW amplitude reached at the dot increases due to the SAW modulation risetime, and therefore the highest probability reached rises.

Figure~\ref{fig4}(b) shows the evolution of the maximum emission probability at $\Delta \text{t}_{\text{max}}= \SI{865}{ns}$ as a function of the SAW amplitude. As previously demonstrated, the SAW amplitude needs to reach a threshold to be able to catch the electron and the injection process shows an activation law with respect to the SAW amplitude. At high SAW amplitude, the injection process has a probability close to unity, and is expected to be adiabatic [Fig.~\ref{fig3}(b)]. Figure~\ref{fig4}(c) shows the evolution of the maximum emission probability at $\Delta \text{t}_{\text{max}}$ as a function of the nanosecond pulse amplitude for two different points P from which the nanosecond pulse is applied. Similarly to the results presented in Fig.~\ref{fig2}(f) for the SAW amplitude, we obtain a threshold behavior in gate voltage. It is worth noting that no broadening of the distribution is observed when the electron is buried deeper into the source dot. This is consistent with the electron staying in the ground state of the source dot potential during the injection process.

In conclusion, we have studied the injection process of a single electron from a static quantum dot into a SAW-induced moving quantum dot. For a specific dot-channel potential gradient at the injection position, the injection probability can be tuned from zero to one by varying the SAW amplitude, pointing at a probabilistic process. At sufficiently large SAW amplitude, only one period of the SAW modulation is required to transfer the electron. In this regime, the injection process is expected to be adiabatic, with the electron remaining, at any time, in the ground orbital state of the trapping potential. Such a regime will be relevant to reduce errors in quantum protocols where single electrons are transferred in moving quantum dots.

\begin{acknowledgments}

We acknowledge technical support from the ``Poles'' of the Institut N\'eel as well as from Pierre Perrier. M.Y. acknowledges financial support by JSPS, Grant-in-Aid for Scientific Research A (No. 26247050) and Grant-in-Aid for Challenging Exploratory Research (No. 25610070). S. Tarucha acknowledges financial support by JSPS, Grant-in-Aid for Scientific Research S (No. 26220710), MEXT KAKENHHI ``Quantum Cybernetics,'' MEXT project for Developing Innovation Systems, and JST Strategic International Cooperative. A.L. and A.D.W. acknowledges DFG via TRR160, the support of the BMBF  Q.com-H  16KIS0109, Mercur  Pr-2013-0001 and the DFH/UFA  CDFA-05-06. B.B.  acknowledges financial support from ``Fondation Nanosciences.'' T.M. acknowledges financial support from ERC ``QSPINMOTION.''

\end{acknowledgments}

\bibliography{biblio}
\end{document}